\documentclass[prl,aps,showpacs,showkeys,twocolumn]{revtex4}
\usepackage{graphicx}
\usepackage{epsfig}

\begin{document}
\title{Origins of elastic properties in ordered nanocomposites}

\author{R. B. Thompson}
\altaffiliation[Present address: ]{Department of Physics, University of 
Waterloo, 200 University Avenue West, Waterloo, Ontario, Canada N2L 3G1}
\author{K. \O. Rasmussen}
\email[Corresponding author: ]{kor@lanl.gov} 
\author{T. Lookman}
\affiliation{Theoretical Division, Los Alamos National Laboratory, Los Alamos, 
New Mexico 87545}
\date{\today}

\begin{abstract}
We predict a diblock copolymer melt in the lamellar phase with added spherical 
nanoparticles that have an affinity for one block to have a 
\emph{lower} tensile modulus than a pure diblock copolymer system. This 
weakening is due to the swelling of the lamellar domain by 
nanoparticles and the displacement of polymer by elastically inert fillers. 
Despite the overall decrease in the tensile modulus of a polydomain  
sample, the shear modulus for a single domain increases dramatically.
\end{abstract}

%83.80.Uv: Block copolymers
%81.05.Qk: Reinforced polymers and polymer-based composites
%62.20.Dc: Elasticity, elastic constants
\maketitle

Polymer nanocomposites are being extensively investigated because of the 
improvement in material properties that result from the addition of nanoscopic 
filler particles to the polymer matrix \cite{Schmidt2003, Balazs2000, Vaia2001, 
Vaia2001b}. In addition to 
their practical importance, such composites offer diverse scientific challenges, 
 combining
ideas from colloid science, polymer physics and chemistry, as well as material 
science. Polymer nanocomposites become even more interesting when the polymer 
matrix 
consists of a block copolymer, capable of self-assembling into a wide range of 
ordered 
nanoscaled structures --- nanoparticles can then be sequestered in 
certain domains to form ordered nanocomposites \cite{Ha2004,Bockstaller2001, 
Bockstaller2003, Jain2002}. The simultaneous amphiphilic and colloidal 
self-assembly  taking
place in such ordered nanocomposites gives them complex structures  
\cite{Thompson2002b} and makes the structure-property relationship particularly 
intriguing. Since there is little understanding of the mechanical 
properties that arise in ordered nanocomposites, we present in this theoretical 
work a first investigation of the \emph{origins} of the elastic properties of 
an ordered nanocomposite with spherical nanofillers. 

Buxton and Balazs \cite{Buxton2003} have studied a phenomenological model of 
nanosphere 
filled block copolymer systems in which a hybrid Cahn-Hilliard/Brownian dynamics 
simulation is used as input to a lattice spring model of the 
elastic moduli. Their approach provides a versatile and useful method of 
predicting properties, but lacks polymeric detail in the elasticity portion 
of the 
simulation. Furthermore, they examine filled block copolymer systems in the 
solid state, where all morphological evolution is disregarded as the 
system is distorted.

We examine the elastic properties of a melt state ordered 
nanocomposite using self-consistent field theory (SCFT).  SCFT 
is a coarse-grained, first principles approach that been successful in 
dealing with block copolymer structure \cite{Matsen2002}. In the framework of 
this theory, local monomer density profiles of different block copolymer 
chemical species are represented self-consistently using chemical 
potential fields. Both the densities and the 
fields are then used to determine the free energy for the system, and if 
desired, the 
internal energies and entropies can be explicitly calculated. SCFT has been 
extended 
to deal with hard nanosphere/block copolymer nanocomposites by the incorporation 
of a density functional theory particle contribution \cite{Thompson2001, 
Balazs2003}. Further, Tyler and Morse have 
demonstrated that the linear elastic behavior of a melt block copolymer system, 
which is 
quasi-statically deformed can be well characterized using SCFT \cite{Tyler2003}. 
We have
recently adapted this approach to an efficient real space, pseudo-spectral 
method \cite{Thompson2004} and found an increasing elastic modulus in multiblock 
copolymer systems as a function of block number, in qualitative agreement with 
experiment \cite{Spontak2001}. Here, we combine these two advances in order not 
only to predict the effect on the elastic properties of adding nanoparticles to 
a block 
copolymer melt, but also to explain the \emph{physical origins} of the observed 
effects.

We will study the prototypical system consisting of a symmetric AB diblock 
copolymer melt in 
the lamellar phase with added spherical nanoparticles that have an affinity for 
the A block of the copolymer. Consequently, a lamellar morphology with the 
particles sequestered in the A phase is 
being considered, and the system's tetragonal symmetry is elastically 
characterized 
by just five independent non-zero components of the elastic modulus 
tensor. Additionally, the system is in a melt 
state so that deformations parallel to the lamellar structure have no effect on 
the 
free energy of the system \cite{Thompson2004}. We are thus left with only two 
relevant moduli, 
$K_{33}$ and $K_{44}$. Therefore we deform the system quasi-statically in two 
ways; it is subjected to an extension/compression, and to a simple shear. These 
deformations allow us to determine the $K_{33}$ and $K_{44}$ components of the 
elastic modulus tensor \cite{Landau1999}, corresponding to 
extension/compression and shear moduli, respectively. These components are found 
by taking the second derivative of the SCFT free energy with respect to the 
relative deformation. Greater detail on the methodology can be found in Ref. 
\cite{Thompson2004}. 

We chose a system with a segregation of $\chi N=25$ between the A and B blocks, 
with the particles considered to be of the A species. $\chi$ is the 
Flory-Huggins monomer segregation parameter and $N$ is the degree of 
polymerization of the entire diblock. The particles radius was chosen to be  
$0.725R_g$, where $R_g$ is the unperturbed radius of gyration of a diblock 
molecule. Finally, the particle-to-diblock volume ratio was $\sim 3.6$, and a 
$15\%$ volume fraction of spherical fillers was added. The system was deformed 
in the two 
ways described above and compared with a neat diblock system similarly deformed. 
The $K_{33}$ and $K_{44}$ moduli in each case were used to find a tensile 
modulus by averaging over a polydomain sample according to the Hill prescription 
\cite{Schreiber1973} as described in Ref. \cite{Thompson2004}. 

For the pure diblock system a dimensionless tensile modulus of $\sim 0.4$ was 
found while 
for the nanocomposite system, the modulus was $\sim 0.3$. Thus we predict that 
the addition of 
nanoparticles will \emph{weaken} the material. Normally, one expects the 
addition of nanofillers to strengthen the composite, 
rather than weaken it \cite{Balazs2000}, although a reduced modulus has been 
observed experimentally in exfoliated layered silicate/triblock nanocomposites 
\cite{Ha2004}. The reduced modulus found for our present system can be explained 
by decomposing 
the $K_{33}$ and $K_{44}$ moduli as shown in tables \ref{tab-K33} and 
\ref{tab-K44}.
\begin{table}
\begin{tabular}{|c|c|c|c|}  \hline
modulus & $0\%$ & $15\%$ & $\Delta$ modulus \\ 
\hline 
$K_{33}$ & 4.67 & 3.40 & 1.27 \\
\hline
$K_{33}^U$ & 2.75 & 2.03 & 0.72 \\
\hline
$K_{33}^{S_T}$ & -0.58 & -0.56 & -0.02 \\
\hline
$K_{33}^{S_A}$ & 1.25 & 0.36 & 0.89 \\
\hline
$K_{33}^{S_B}$ & 1.25 & 1.39 & -0.13 \\
\hline
$K_{33}^{S_{id}}$ & 0.0 & 0.22 & -0.22 \\
\hline
$K_{33}^{S_{st}}$ & 0.0 & -0.04 & 0.04 \\
\hline
\end{tabular}
\caption{Components of the $K_{33}$ elastic modulus for $0\%$ and $15\%$ added 
nanoparticles. The change in the modulus between the pure diblock and the filled 
diblock system is recorded in the last column.}
\label{tab-K33}
\end{table}
\begin{table}
\begin{tabular}{|c|c|c|c|}  \hline
modulus & $0\%$ & $15\%$ & $\Delta$ modulus \\ 
\hline 
$K_{44}$ & 0.01 & 0.05 & -0.04 \\
\hline
$K_{44}^U$ & 1.10 & 0.94 & 0.16 \\
\hline
$K_{44}^{S_T}$ & -0.53 & -0.51 & -0.02 \\
\hline
$K_{44}^{S_A}$ & -0.28 & -0.22 & -0.06 \\
\hline
$K_{44}^{S_B}$ & -0.29 & -0.07 & -0.22 \\
\hline
$K_{44}^{S_{id}}$ & 0.0 & -0.02 & 0.02 \\
\hline
$K_{44}^{S_{st}}$ & 0.0 & -0.07 & 0.07 \\
\hline
\end{tabular}
\caption{Components of the $K_{44}$ elastic modulus for $0\%$ and $15\%$ added 
nanoparticles. The change in modulus between the pure diblock and the filled 
diblock system is recorded in the last column.}
\label{tab-K44}
\end{table} 
These tables show total $K_{33}$ and $K_{44}$ moduli values, respectively, for 
$0\%$ and $15\%$ added fillers, as well as the difference between the filled and 
unfilled system moduli. The tables also give the contributions to the moduli of 
the internal energy ($U$), translational entropy ($S_T$), and the A and B 
conformational 
entropies of the diblocks ($S_A$ and $S_B$). Also listed are the ideal gas 
($S_{id}$) and steric ($S_{st}$) 
contributions of the filler particles. For information on how these components 
are calculated, see Ref. \cite{Thompson2004}. Comparing the total $K_{33}$ and 
$K_{44}$, it is observed that the main contribution to the tensile modulus 
arises 
from the extensional modulus $K_{33}$. Furthermore, the main drop in $K_{33}$ 
from the $0\%$ to the $15\%$ case is a result of the internal energy 
contribution and the A block conformational entropy contribution.

\begin{figure}
\begin{center}
\scalebox{0.4}{\includegraphics{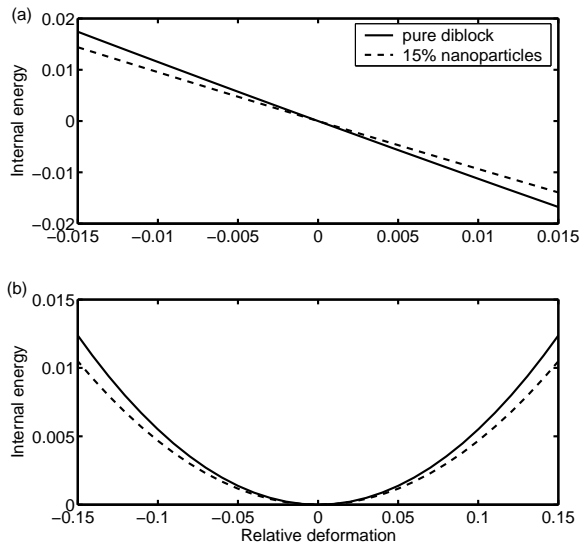}}
\end{center}
\caption{Internal energy contribution to the free energy versus 
relative distortion for extension/compression (a) and shear (b) of a diblock 
system. 
The neat diblock free energy is plotted with the solid line whereas the $15\%$ 
filled system is shown with dashed line. In (a), negative deformations are 
compressions while positive deformations are extensions. In all plots, energies 
have been zeroed around the equilibrium spacing, which is represented by 
$\epsilon=0$.}
\label{fig-Ucontr}
\end{figure}

These two contributions can be examined separately. Upon 
extension/compression, SCFT shows that the lamellar interfacial width remains  
practically unchanged 
in both the filled and unfilled cases. Furthermore, the A and B domains are well 
segregated both before and after 
deformation, indicating that there are few A monomers in the B region and 
\emph{vice 
versa}. Thus the absolute amount of energetically unfavorable AB contacts is the 
same before 
and after deformation, whereas the domain size changes. Adapting an expression 
of 
Matsen and Bates 
\cite{Matsen1997}, the internal energy contribution to the free energy is
\begin{equation}
\frac{U}{k_BT} \propto \frac{1}{V} \int d{\bf r} \left[\varphi_A({\bf 
r})+\varphi_p({\bf r}) \right]
\varphi_B({\bf r})    \label{U1}
\end{equation}
where $T$ is the temperature and $k_B$ is Boltzmann's constant. $\varphi_A({\bf 
r})$, $\varphi_B({\bf r})$ and $\varphi_p({\bf r})$ are the local volume 
fractions of A and B monomers, and particles, respectively. The integral in 
Eq. (\ref{U1}) is constant under 
the conditions described above, so that the internal energy is inversely 
proportional to the volume. The lamellar morphology is one dimensional, so that 
the internal energy can be written as inversely proportional to the equilibrium 
domain size $d^*$.
\begin{equation}
\frac{U}{k_BT} = \frac{\alpha}{d^*}  \label{U2}
\end{equation}
where $\alpha$ is a constant. A relative deformation $\epsilon$ that leaves 
the interfacial width and the bulk mixing unchanged, changes Eq. (\ref{U2}) into
\begin{equation}
\frac{U}{k_BT} = \frac{\alpha}{d^* (1+\epsilon)}  .   \label{U3}
\end{equation}
The SCFT internal energy for extensions/compressions is shown in Fig. 
\ref{fig-Ucontr}(a).
Both for pure diblock and for filled systems, the behavior reflected in Eq. 
(\ref{U3}) is 
observed. An estimate of the modulus $K_{33}^U$ can be found by taking the  
second derivative of Eq. (\ref{U3}) with respect to the relative deformation, 
which 
gives
\begin{equation}
K_{33}^U = \left.\frac{d^2 (U/k_BT)}{d \epsilon^2}\right|_{\epsilon=0} = 
\frac{2\alpha}{d^*}   \label{K33U}
\end{equation}
showing that the $U$ contribution to the $K_{33}$ modulus is also inversely 
proportional to the equilibrium spacing. SCFT shows that the addition of filler 
particles enlarges the equilibrium domain size of the nanocomposite compared
to the pure diblock system. Equation (\ref{K33U}) then indicates that the 
modulus will drop, as observed. In other words, the modulus is weakened 
partially because there is less interface per volume in the nanocomposite 
compared to the pure diblock.

\begin{figure}
\begin{center}
\scalebox{0.4}{\includegraphics{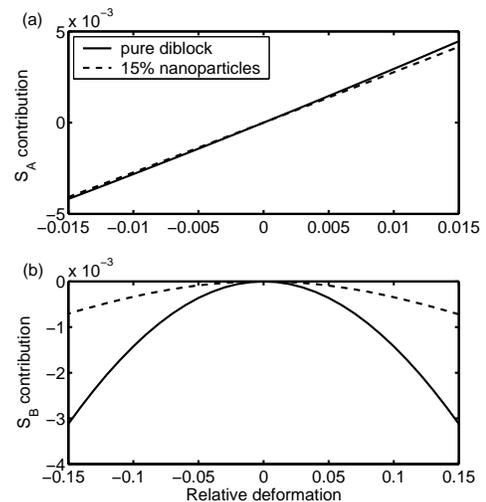}}
\end{center}
\caption{Conformational entropy contribution to the free energy versus 
distortion for extension/compression (a) and shear (b) of a diblock system. 
The neat diblock free energy is plotted with the solid lines whereas the $15\%$ 
filled system is shown with dashed lines. In (a), negative deformations are 
compressions while positive deformations are extensions; the $S_A$ contribution 
is shown in (a). The $S_B$ contribution is shown in (b). In all plots, energies 
have been zeroed around the equilibrium spacing, which is represented by 
$\epsilon=0$.}
\label{fig-Scontr}
\end{figure}

The A block configurational entropy also contributes to the overall drop in 
modulus. The filler particles have no 
configurational entropy; in the filled system the $S_A$ energy portion will rise 
(drop) under extension (compression) at a slower rate because there is a smaller 
fraction of chains to stretch 
(relax). This can be seen in Fig. \ref{fig-Scontr}(a). 
Thus filler particles weaken the material because they displace polymers that 
have  
stretching energy, which could contribute positively to the elastic modulus. In 
other words, the modulus is weakened partially because there is less 
polymer and more elastically inert filler per volume.

As mentioned, the shear modulus $K_{44}$ only makes a small contribution 
to the tensile modulus, but it is interesting to examine it nonetheless. Table 
\ref{tab-K44} shows that while the $K_{33}$ modulus drops upon the addition of 
fillers, $K_{44}$ \emph{increases} by more than 
$400\%$. From table \ref{tab-K44}, this increase can be seen to be due to a very 
large increase in the B block conformational entropy contribution to the 
modulus. This is partially reduced by a drop in the internal energy 
contribution, but is still very large.

As in the extension/compression case, the interfacial profiles do not change 
significantly upon shearing of the sample. The consequence is that the 
internal energy again is  
inversely proportional to the volume of the system, as previously explained. The 
B block conformational entropy contribution to the free energy behaves 
similarly; from Matsen and Bates \cite{Matsen1997}, the B block energy is
\begin{equation}
\frac{-S_B}{k_B} \propto -\frac{1}{V} \int d{\bf r} \left\{\rho_J({\bf r})\ln 
q^\dag ({\bf r},f)+w_B({\bf r}) \varphi_B({\bf r}) \right\}  \label{SB1}
\end{equation}
where $\rho_J({\bf r})$ is the distribution of diblock junction points, 
$w_B({\bf r})$ is the chemical potential field for the B monomer distribution, 
and $q^\dag ({\bf r},s)$ is a SCFT propagator. A detailed explanation of Eq. 
(\ref{SB1}) can be found elsewhere \cite{Matsen1997, Matsen2002}. If we for the 
moment ignore the contribution of the integral in Eq. (\ref{SB1}), the B 
conformational contribution can be written as 
\begin{equation}
\frac{-S_B}{k_B} = -\frac{\beta}{V \sin\theta}   \label{SB2}
\end{equation}
where $\beta$ is a constant. A $\sin\theta$ has been added to the denominator of 
Eq. (\ref{SB2}) to account for 
the effect of a shear change in volume. When the sample is sheared, the volume 
is 
reduced; $\theta$ is a measure of the amount of shear with $\theta=\pi/2$ 
representing a non-sheared system. See Ref. \cite{Thompson2004} for more 
explanations. The internal energy (\ref{U2}) will be similarly affected. From 
Ref. \cite{Thompson2004}, the shear angle is related to the relative 
distortion through $\epsilon=\cot \theta$. The internal energy and B 
conformational entropy are then
\begin{eqnarray}
\frac{U}{k_BT} &=& \frac{\alpha}{V} \sqrt{\epsilon^2+1},  \label{U4}  \\
\frac{-S_B}{k_B} &=& -\frac{\beta}{V} \sqrt{\epsilon^2+1} . \label{SB4} 
\end{eqnarray}
Figures \ref{fig-Ucontr}(b) and \ref{fig-Scontr}(b) show that the internal 
energy and B block contribution to the free energy obey the relationships 
(\ref{U4}) and (\ref{SB4}), respectively, for both the filled and unfilled 
systems. The moduli for these are found through the second derivative with 
respect to $\epsilon$ and are
\begin{eqnarray}
K_{44}^U &=& \frac{\alpha}{V},  \label{U5}  \\
K_{44}^{S_B} &=& -\frac{\beta}{V}  . \label{SB5}
\end{eqnarray}
Both $K_{44}^U$ and $K_{44}^{S_B}$ are inversely proportional to the equilibrium 
volume, with the important difference of a minus sign. Thus as the equilibrium 
volume is enlarged upon the addition of fillers, the $S_B$ contribution to the 
modulus becomes a smaller \emph{negative} number, so the material is 
\emph{stronger}. This effect is somewhat reduced as the $U$ contribution to 
the modulus will become smaller for the larger equilibrium volume of the filled 
system. In other words, diblock molecules' entropies want to help shear the 
system (see Fig. \ref{fig-Scontr}(b)).  The larger domain size of the filled 
system means that there is less stretching energy per volume (see eq. 
(\ref{SB4})), the molecules are 
more relaxed to begin with at equilibrium, 
and so are less inclined to help deform the system, which makes the material 
stronger. This effect is reduced since there is less interface per volume in the 
filled system. The A block is not treated the same way, since with the fillers 
added the $S_A$ contribution to the free energy cannot be written in a form 
such as Eq.(\ref{SB2}).

In summary, we have calculated the tensile modulus for a neat diblock copolymer 
system and for a diblock nanocomposite with $15\%$ added nanospheres. Both 
systems were considered to be in the lamellar phase, and deformations were 
applied quasi-statically. The elastic modulus tensor components $K_{33}$ and 
$K_{44}$ were found and used to derive a tensile modulus for a polydomain  
sample. It was found that the addition of nanoparticles \emph{weakened} the 
material. This was attributed to the larger lamellar domain size of the 
equilibrium filled system --- it had less interface per volume --- and to the 
displacement of polymer by the filler particles --- there was less elastic 
polymer per volume. Although, the shear modulus contributed little, it was acted
on by similar mechanisms, with the result that it \emph{increased} dramatically 
upon the addition of nanofillers. Given that ordered nanocomposites of the sort  
described here can now be realized, we believe our predictions should be 
amenable to experimental verification. It would be interesting to change the 
distribution of the particles in the block copolymer through changes of wetting 
properties, particle size, or volume fraction. Such changes in structure could 
significantly change the properties. 

\begin{acknowledgments}
Work at the Los Alamos National Laboratory was performed under the auspices 
(contract W-7405-ENG-36) of the U.S. Department of Energy. The authors are  
grateful to 
Yung-Hoon Ha \textit{et al.} for providing a copy of their manuscript prior to 
publication.
\end{acknowledgments}

\end{document}